\documentclass[aps,reprint, showpacs, epsfig]{revtex4-1}
\usepackage{graphicx}
\usepackage{subfigure}
\usepackage{epsfig}
\usepackage{latexsym}
\usepackage{epstopdf} 
\usepackage{color}
\usepackage{mathtools}

\usepackage{hyphenat}
\let\oldref\ref
\renewcommand{\ref}[1]{(\oldref{#1})}
\usepackage[colorlinks=true,linkcolor=blue,urlcolor=blue,citecolor=blue]{hyperref}
\usepackage{amsmath}

\begin{document}
\title{Super-Arrhenius behaviour of molecular glass formers}
\author{Ankit Singh and Yashwant Singh}
\affiliation{Department of Physics, Banaras Hindu University,
Varanasi-221 005, India.}
\date{\today}

\begin{abstract}
A theory is developed to calculate values of the potential energy
barriers to structural relaxation in molecular glass formers from
the data of static pair correlation function.
The barrier height is shown to increase due to increase in number
of the ``stable bonds'' a particle forms with its neighbours and 
energy of each bond as liquids move deeper into 
the supercooled (supercompressed) region. We present results 
for a system of hard-spheres and compare calculated 
values of the structural relaxation time with experimental and 
simulation results.
\end{abstract}


\maketitle

The structural relaxation time of a molecular glass former grows by many 
orders of magnitude over a small range of temperatures when the system 
is cooled close to the glass transition temperature \cite{Angell}.
The glass transition is linked to dynamical arrest caused by particles being 
trapped in cages formed by their nearest neighbours \cite{Kegel,Brown}.
It is widely accepted that the dynamics close to the glass transition is dominated 
by activation \cite{Goldstein}. If the potential energy barriers to relaxation were constant
in temperatures, the relaxation time would follow the Arrhenius law.
The super-Arrhenius behaviour suggests that the potential energy barriers
in molecular glass formers increase with decreasing temperature and increasing density.
The ubiquity of the phenomenon, irrespective of molecular details points
to a collective or cooperative behaviour characterized by a length scale
that grows as one approaches to the glass transition.
Beginning at least from Adam and Gibbs \cite{Adam}
who introduced the concept of ``cooperatively rearranging regions'' in the
mid - 1960'$^{s}$, many microscopic models \cite{R,Ritort,Kirkpatrick,Chandler,
Berthier,Whitelam} have been developed to uncover the physical
mechanism behind growth of the cooperative length scale.
One of the issues has been to define and determine objectively such a
length scale \cite{Berthier,Whitelam,Ediger}, and relate it with the
potential energy barrier.

In dealing with classical many-body particle systems one often integrates out 
kinetic energy of particles and considers only potential energy of interactions
in framing a theory or in simulations. In this letter we show that when kinetic
energy is allowed to compete with effective potential energy felt by particles
in a system, a new way of understanding properties of dense systems emerges.
Such an idea was first proposed by Hill \cite{Hill} and used by Stogryn and 
Hirschfelder \cite{Stogryn} and others \cite{ysingh} to describe equilibrium and transport
properties of gases.

A particle in a dense system feels potential energy barrier created by its 
neighbours. Depending upon height of the barrier and the relative momenta 
of surrounding particles, the central particle may get trapped and bonded
(defined below) with neighbouring particles. A particle whose total energy
is higher than the barrier moves freely and
collides with other particles. The concentration of these particles depends
on density and temperature; the potential barrier becomes higher on increasing 
the density and lowering the temperature and kinetic energy of particles 
decreases on decreasing the temperature. A
molecular liquid at high densities and low temperatures can therefore be considered
as a network of particles connected with each other by (non-chemical) bonds
with some free particles which move around and collide with other particles.
Depending upon bonding energies, the life time of bonds may vary from 
microscopic to macroscopic time. When a particle dissociates from the network
either by collision or by thermal activation, it may initiate breaking of 
neighbouring bonds and creating a dynamical active domain \cite{Harrowell,Garrahan}.
The precipitous onset of slowness can be associated with increasing number
of bonds and the larger bond energy with which particles are bonded with neighbours.

One way to find number of bonds formed by a particle with its neighbours 
is to use data of static pair correlation function. The theory we 
describe is applicable to all those systems for whose values of pair 
correlation function in supercooled (super-compressed) region are 
available. Here we consider a system of hard spheres and use data of the
radial distribution function (RDF) evaluated from an approximate integral
equation theory \cite{Singh} (our aim here is to show usefulness of the theory
rather than numerical rigour).

In case of a system of hard-spheres where potential is zero when particles 
do not overlap and infinite otherwise, temperature becomes irrelevant 
apart from rescaling quantities, the natural control parameter is the 
packing fraction $\eta=\frac{\pi}{6}\rho{\sigma}^3$, where $\rho$
is number density and $\sigma$ particles diameter.
Experimentally, hard-sphere systems are obtained using colloidal
particles \cite{Pusey}, emulsions, or granular materials \cite{Liu}.
The fluid-crystal transition takes place at $\eta=0.494$ and the melting transition 
at $\eta=0.545$. When the system is compressed following a protocol 
which avoids crystallization, the structural relaxation time $\tau_{\alpha}$
increases rapidly showing super-Arrhenius behaviour. Whether $\tau_{\alpha}$ 
diverges at a density lower than the random close packed density , $\eta_{rcp}(\simeq0.64)$ 
or not is still a highly debated
issue \cite{Donev,Martinez,Parisi}. Kinetic arrest must occur at $\eta_{rcp}$
because all particles block each other at that density.

    The RDF g(r) of a homogeneous and isotropic system consisting of particles 
of mass m in the centre of mass coordinates can be written as 

\begin{equation}
 g(r)=\left(\frac{\beta}{2\pi\mu}\right)^{\frac{3}{2}}\int d{\bf p} \  e^{-\beta (\frac{p^2}{2\mu} + W(r))} , 
\end{equation}
 
where $\beta=(k_{B}T)^{-1}$ is the inverse temperature measured in unit of
the Boltzmann constant $k_{B}$ and ${\bf p}$  is the relative momentum of 
a particle of mass $\mu=m/2$. The effective potential $W(r)=-k_{B}T\ln g(r)$ 
\cite{Hansen} is sum of the (bare) pair potential and the system induced potential
energy of interaction between a pair of particles separated by distance r.
In Fig. 1 we plot $\beta W(r)$ for a system of hard spheres at 
$\eta=0.59$ as a function of r expressed in unit of $\sigma$.
The curve has several maxima and minima. We denote a region between two 
maxima i-1 and i $(i \geq 1)$ as $i^{th}$ shell and the minimum of the shell by
$\beta W_{di}$. The value of $i^{th}$ maximum is denoted by $\beta W_{ui}$
and its location by $r_{hi}$.

\begin{figure}[t]
\includegraphics[scale=0.4]{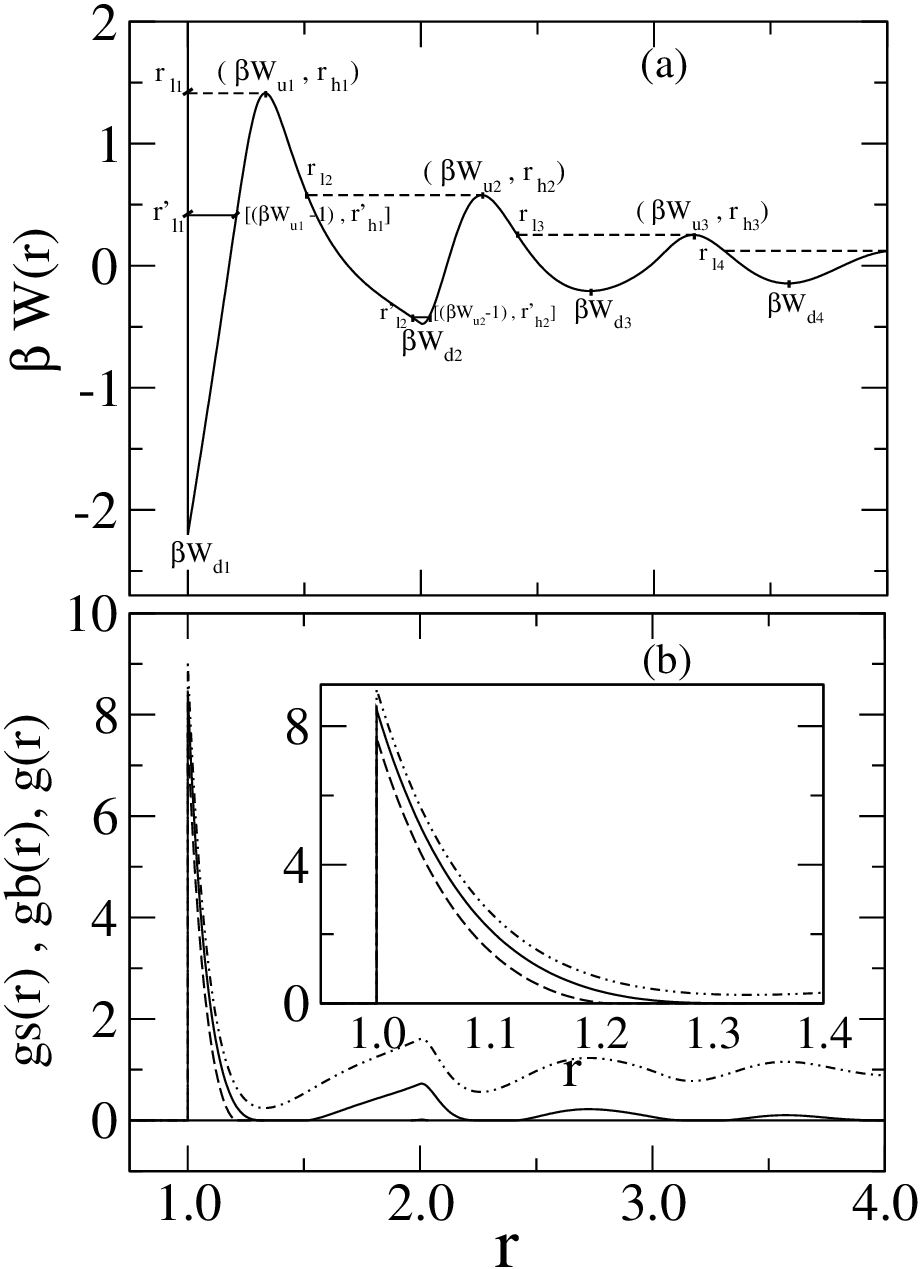}

\caption{{(a). The reduced effective potential $\beta W(r)$ between a pair of particles separated by distance r
(expressed in unit of hard-spheres diameter $\sigma$) in a system of hard-spheres at a packing fraction
$\eta=0.59$. $\beta W_{ui}, r_{hi}$ are, respectively, value and location of $i^{th}$ maximum and
$r_{li}$ is the location on the left hand side of the shell where $\beta W_{i}(r)=\beta W_{ui}$ (shown by dashed line).
The location $r'_{li}$ and $r'_{hi}$ are values of r on the left and the right hand side of the shell
where $\beta W_{i}(r)=\beta W_{ui}-1$ (shown by full line). $\beta W_{di}$ is the depth of the $i^{th}$ shell.}
\\
{(b). Radial distribution functions g(r) (dash-dotted line), $g_{b}(r)$ (full line) and $g_{s}(r)$ (dashed line)
vs r at $\eta=0.59$. The various peaks correspond to various shells around a central particle. While g(r)
oscillates around one, $g_{b}(r)$ and $g_{s}(r)$ become zero at boundaries defined by ($r_{li}, r_{hi}$)
for $g_{b}(r)$ and ($r'_{li}, r'_{hi}$) for $g_{s}(r)$. In the inset we show how values in the first shell
differ from each other.}}

%vspace{-1cm}
\label{fig-1}

\end{figure}

All particles of $i^{th}$ shell whose energies are less or equal to $\beta W_{ui}$
i.e; $\beta(\frac{p^2}{2\mu}+W_{i}(r))\leq \beta W_{ui}$ get confined in the shell
and can be considered to be bonded with the central particle.
The contribution made to g(r) by these particles is
\begin{equation}
g_{bi}(r)=4\pi(\frac{\beta}{2\pi\mu})^{\frac{3}{2}}  
e^{-\beta W_{i}(r)} \int_{0}^{\sqrt{2\mu(W_{ui}-W_{i}(r))}}  e^{-\frac{\beta p^2}{2\mu}} p^2 dp \nonumber
\end{equation}

\begin{equation} 
=e^{-\beta W_{i}(r)} \  \frac{\Gamma(\frac{3}{2},\beta (W_{ui}-W_{i}(r)))}{\Gamma(\frac{3}{2})}  , 
\end{equation}

where $\Gamma(m,n)$ is the incomplete gamma function and $W_{i}(r)$ is the 
effective potential of $i^{th}$ shell in the range $r_{li}\leq r \leq r_{hi}$.
Here $r_{li}$ is the value of r where $\beta W_{i}(r_{li})= \beta W_{ui}$
on the left hand side of the shell (see Fig. 1). In Fig. 1(b) we plot g(r) and 
$g_{b}(r)$ as a function of r at $\eta=0.59$.
The number of bonded particles of $i^{th}$ shell at packing fraction $\eta$
is

\begin{equation}
n_{bi}(\eta) = 24 \eta \int_{r_{li}}^{r_{hi}} g_{bi}(r) r^2 dr .
\end{equation}

The total number of particles bonded with the central particle is
$N_{b}(\eta) = \sum_{i} n_{bi}(\eta)$.
As shown in Fig. 2(a) by full line, $N_{b}$ increases rapidly
above the freezing density. This is due to increase in number of shells that 
surround the central particle and values of $\beta W_{ui}$ and $\beta W_{di}$
with $\eta$. It may, however be noted that these particles (or bonds) are embedded 
in a system which is equipped with thermal energy $k_{B}T$.
Therefore, all those particles whose energies lie between $\beta W_{ui}-1$
and $\beta W_{ui}$ may not remain bonded for long due to thermal
fluctuations; the life time depends on the bonding energy. 
We call these particles meta-stably bonded (henceforth referred to as m-particles)
and those particles whose energies lie between $\beta W_{di}$ and $\beta W_{ui}-1$, 
stably bonded (henceforth referred to as s-particles).
Most particles of $N_{b}$ shown in Fig. 2(a) surrounding the central
particle over a length of the pair correlation function are m-particles
with bond energies much smaller than the thermal energy and are therefore
transient.

The contribution made to g(r) by the s-particles of $i^{th}$ shell is 

\begin{equation}
g_{si}(r)=4\pi(\frac{\beta}{2\pi\mu})^{\frac{3}{2}}  
e^{-\beta W_{i}(r)} \int_{0}^{\sqrt{2\mu(W_{ui}-K_{B}T-W_{i}(r))}}  e^{-\frac{\beta p^2}{2\mu}} p^2 dp ,
\end{equation}

where $W_{i}(r)$ is in the range, $r'_{li} \leq r \leq r'_{hi}$. Here $r'_{li}$ and
$r'_{hi}$ are respectively, values of r on the left and the right hand side
of the shell where $\beta W_{i}(r)=\beta W_{ui}-1$ (see Fig. 1).
We show values of $g_{s}(r)$ vs r in Fig. 1(b) by dashed line for $\eta=0.59$.
The number of s-particles with which central particle is bonded, is

\begin{equation}
N_{s}(\eta)=\sum_{i} n_{si}(\eta) \ ; \  \  \  n_{si}(\eta)=24\eta \int_{r'_{li}}^{r'_{hi}} g_{si}(r) r^2 dr .
\end{equation}

We plot, $N_{s}$ vs $\eta$ in Fig. 2(a) along with $N_{b}$ and $N_{m}=(N_{b}-N_{s})$.

\begin{figure}[t]
\includegraphics[scale=0.40]{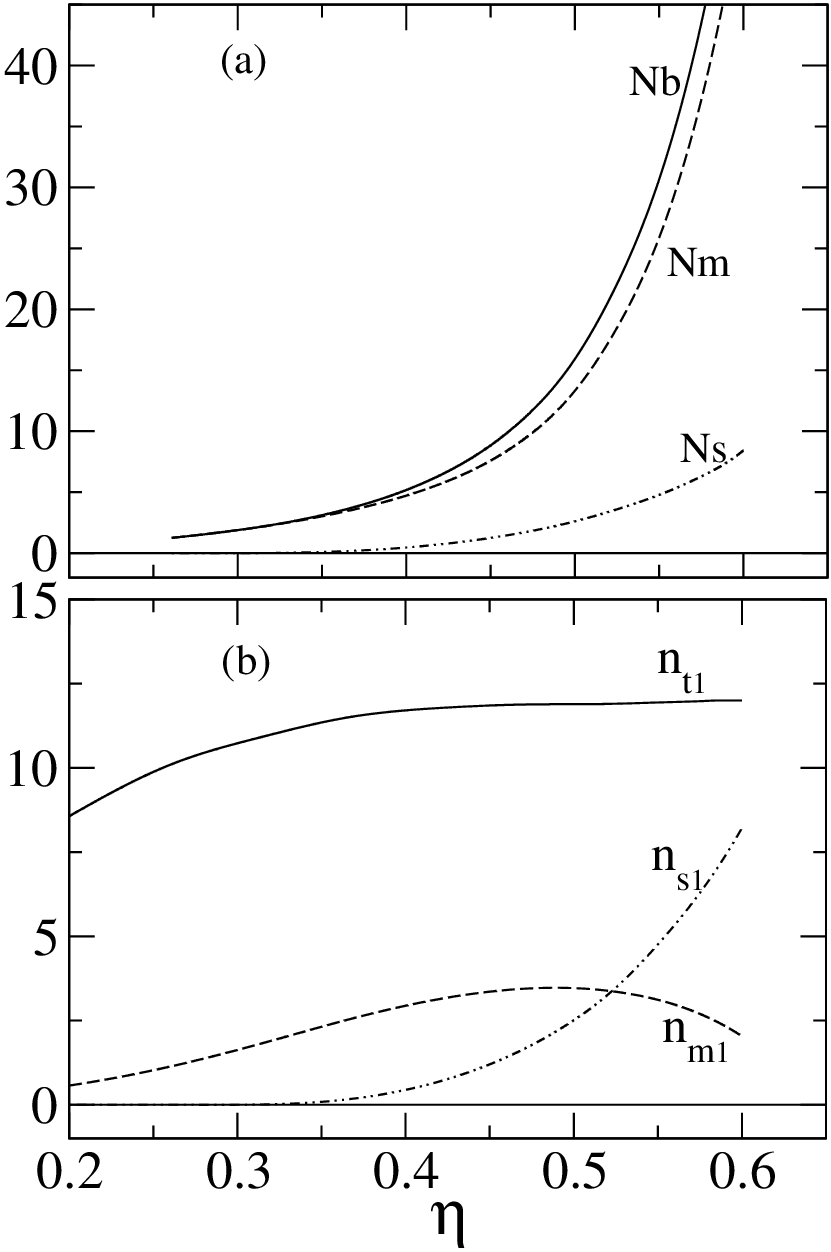}
\caption{{(a). Number of total bonds $N_{b}$, metastable bonds $N_{m}$ and stable bonds $N_{s}$ formed by
a particle in a system of hard-spheres vs packing fraction $\eta$.}
\\
{(b). Number of total particles $n_{t1}$, metastably bonded particles (m-particles) $n_{m1}$
and stably bonded particles (s-particles) $n_{s1}$ in the first shell.
The number $n_{s1}$ increases rapidly and crosses $n_{m1}$ at $\eta \simeq 0.524$.
At $\eta=0.524$ the crossover from nonactivated to activated dynamics takes place,
due to formation of cage by s-particles.}}

%\vspace{-1cm}
\label{fig-2}
\end{figure}

\begin{figure}[]
\includegraphics[scale=0.41]{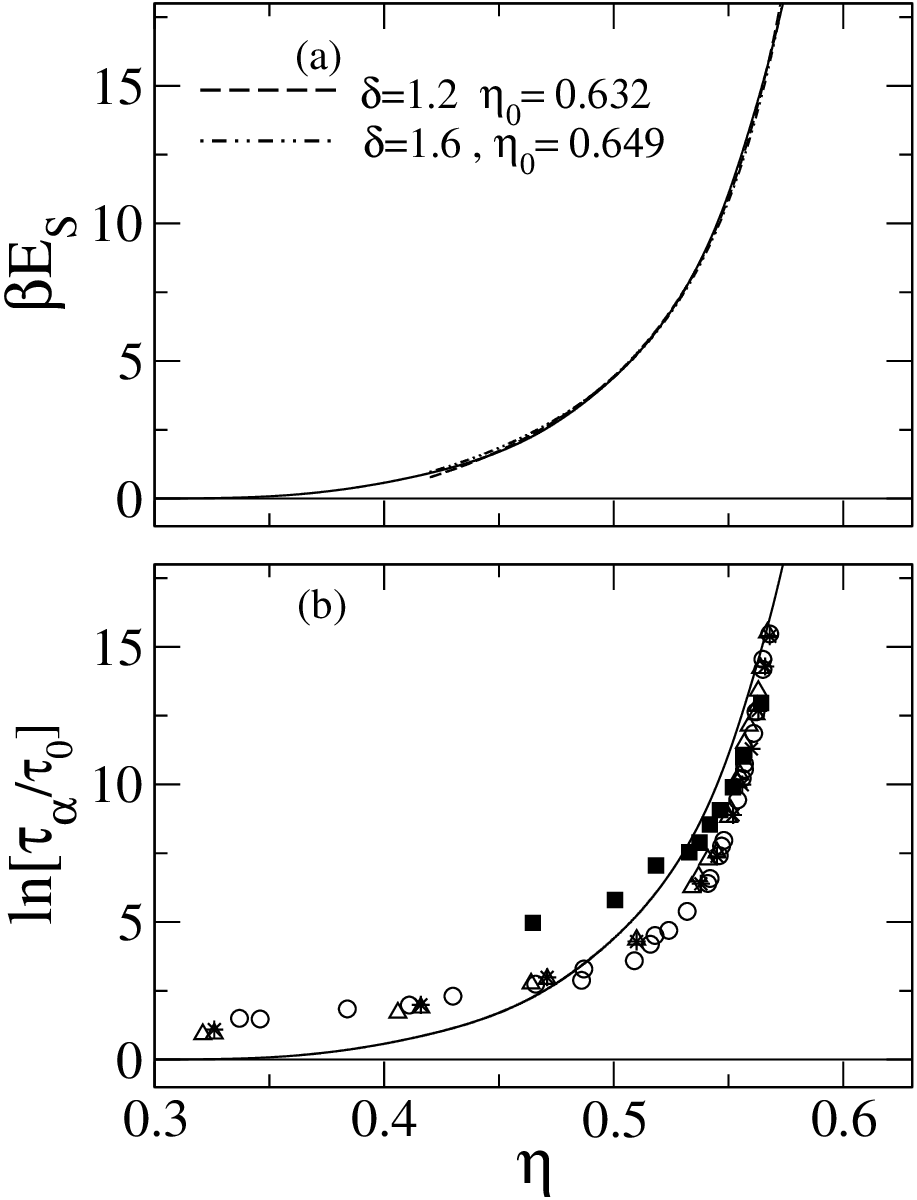}
\caption{{(a). The potential energy barrier (activation energy) $\beta E_{s}$ vs $\eta$.
Values found from the expression $\beta E_{s}(\eta)= A + \dfrac{B}{(\eta_{0}-\eta)^{\delta}}.$
with (i) $\delta=1.2, \eta_{0}=0.632$ (dashed line) and
(ii) $\delta=1.6, \eta_{0}=0.649$ (dash-dotted line) are compared with the calculated values.}
\\
{(b). Calculated values (solid line) of $\ln [\dfrac{\tau_{\alpha}}{\tau_{0}}]$ are compared with experimental values (filled square \cite{Megen}
and open circles \cite{Brambilla}) and simulation values (open triangles \cite{Brambilla} and stars \cite{Witten}).
The values of refs. \cite{Brambilla} and \cite{Witten} are shifted to lower density by an
amount $\Delta \eta=0.03$.}}

%\vspace{-1cm}
\label{fig-3}
\end{figure}

To understand why above a certain density a particle gets trapped by a stiff
barrier and there is a crossover from nonactivated to activated dynamics,
we examine the nature of particles of the first shell surrounding the central
particle as function of $\eta$. In Fig. 2(b) we plot number of total particles
$n_{t1}$, m-particles $n_{m1}$ and s-particles $n_{s1}$.
We note that $n_{t1}$ reaches the maximum value 12 at a density lower than the freezing
density where most particles are still free.
The number of m-particles $n_{m1}$ first increases and after reaching a maximum
value ($\simeq3$) at $\eta \simeq 0.50$ starts decreasing and crosses $n_{s1}$
at $\eta \simeq 0.524$, The number of s-particles ($n_{s1}$) which build up
the potential energy barrier increases rapidly on increasing the density.
We therefore consider $\eta=0.524$ as the density which separates the two
distinct dynamical domains. As for $\eta<0.524$ the potential barrier is inconsequential
 the activation is not 
main mechanism of relaxation; the dynamics can be described by the mode coupling
theory (MCT) \cite{Gotze}. The activated dynamics becomes
dominant for $\eta \geq 0.524$ when the central particle gets surrounded by increasing 
number of s-particles and the barrier starts caging the particle.

The potential energy barrier to relaxation (activation energy) is assumed
to be equal to the energy with which a particle is bonded with s-particles.
Thus the activation energy,

\begin{equation}
\beta E_{s} (\eta)=24\eta \sum_{i}\int_{r'_{li}}^{r'_{hi}} (\beta W_{ui}-\beta W_{i}(r)) g_{si}(r) r^2 dr .
\end{equation}

Here the bonding energy of each bond is measured from the barrier height. 

In Fig. 3(a) we plot $\beta E_{s}(\eta)$ vs $\eta$ and
note that $\beta E_{s}$ increases sharply for $\eta>0.524$.
The energy $\beta E_{s}$ can be considered as the activation energy in the 
Arrhenius law, $\tau_{\alpha}(\eta)=\tau_{0} exp(\beta E_{s}(\eta))$
where $\tau_{0}$ is a microscopic time scale. Out of different functional
forms used to fit the data of $\beta E_{s}(\eta)$, the best fit was found
for $\beta E_{s}(\eta)=A+\frac{B}{(\eta_{0}-\eta)^{\delta}}$.
In Fig. 3(a) we compare values found with (i) $\delta=1.2, \eta_{0}=0.632$ and
(ii) $\delta=1.6, \eta_{0}=0.649$;
though both sets give equally good fit but while one set gives value of $\eta$
where $\tau_{\alpha}$ diverges, lower, the other higher than $\eta_{rcp}$ $(= 0.64)$ 
indicating limitation of such fitting. We emphasize that the fit shown in
Fig. 3(a) not necessarily favours the Vogel-Fulcher-Tammann (VFT) law over other
laws of relaxation as the low density data where nonactivated dynamics mainly
contributes to relaxations are not included in the fitting.

%\begin{figure}[]
%\includegraphics[scale=0.4]{Fig3.eps}
%\includegraphics[width=0.8\textwidth, angle=270]{Fig3.eps}
%\caption{{(a). The potential energy barrier (activation energy) $\beta E_{s}$ vs $\eta$.
%Values found from the expression $\beta E_{s}(\eta)= A + \dfrac{B}{(\eta_{0}-\eta)^{\delta}}.$
%with (i) $\delta=1.2, \eta_{0}=0.632$ (dashed line) and
%(ii) $\delta=1.6, \eta_{0}=0.649$ (dash-dotted line) are compared with the calculated values.}
%\\
%{(b). Calculated values (solid line) of $\ln [\dfrac{\tau_{\alpha}}{\tau_{0}}]$ are compared with experimental values (filled square \cite{Megen}
%and open circles \cite{Brambilla}) and simulation values (open triangles \cite{Brambilla} and stars \cite{Witten}).
%The values of refs. \cite{Brambilla} and \cite{Witten} are shifted to lower density by an
%amount $\Delta \eta=0.03$.}}

%\vspace{-1cm}
%\label{fig-3}
%\end{figure}

In Fig. 3(b) we compare our results of $\tau_{\alpha}$ with experimental
results reported for colloidal hard spheres in refs. \cite{Megen}, \cite{Brambilla} and
simulation results reported in ref. \cite{Witten}. It may, however, be noted that
while our result is for a monodisperse  system, the simulation result \cite{Witten} is for a
50:50 binary mixture with diameters $\sigma$ and $1.4\sigma$ and the experimental
results are for polydisperse systems with polydispersity, s, of about 6\% in \cite{Megen}
and above 10\% in \cite{Brambilla}. From simulation studies \cite{Zaccarelli,PNPusey,EZaccarelli} it has been found that
while moderately disperse hard spheres $(s \sim 5-6\%)$ behave almost like a
monodisperse system, systems with larger dispersity $(s \gtrsim 10\%)$
behave in a complex way. One such effect is to move the glass transition to
higher $\eta$.
The unusual aging behaviour due to strong decoupling between small and large
spheres for $\eta>0.59$ has also been observed \cite{EZaccarelli,Heckendorf}.
The experimental \cite{Brambilla} and simulation \cite{Brambilla,Witten} data plotted in Fig. 3(b) are
shifted to lower density by an amount $\Delta \eta=0.03$.
whereas the experimental values taken from ref. \cite{Megen} are plotted
(shown by filled square) without any shift.
It may be noted that while the shifted values of refs. \cite{Brambilla}
and \cite{Witten} are in good agreement with values of ref. \cite{Megen} for
$\eta>0.53$, considerable difference remains in their values for $\eta<0.53$.
This suggests the need for more experimental data of moderately disperse systems \cite{Segre}.
The theoretical values shown by full line in the figure is in good agreement
with these data for $\eta \gtrsim 0.50$; agreement for $\eta \lesssim 0.50$ is not expected as dynamics in this region as argued above, is other than activation which
has not been considered.

In summary: We developed a theory to calculate the potential energy barriers to
structural relaxation in a molecular glass formers
from the data of static pair correlation function.
A particle in a molecular liquid in the supercooled (supercompressed) region
gets localized by forming (nonchemical) ``stable bonds'' with neighbouring particles.
The number of bonds and the bonding energy increase on lowering the temperature and
increasing the density.
The barrier height (activation energy) is equal to the energy $\beta E_{s}$ with
which a particle is bonded with the s-particles.
When $\beta E_{s}$ is substituted in the Arrhenius law, a super-Arrhenius feature emerges.
Using values of the radial distribution function for a system of hard spheres
found from an approximate integral equation theory \cite{Singh} we calculated the
activation energy. The calculated values of $\tau_{\alpha}$ is found to be in
agreement with the experimental and simulation data in the region where
activated dynamics is dominant.

We acknowledge financial help from the Council of Scientific and Industrial 
Research and the Indian National Science Academy, New Delhi.

\end{document}